\documentclass[journal,article,accept,moreauthors,pdftex,12pt,a4paper,galaxies]{mdpi}
\lastpage{71}
\doinum{10.3390/galaxies3010053}
\pubvolume{3}
\pubyear{2015}
\history{Received: xx / Accepted: xx / Published: xx}

\Title{Thermodynamic relations for entropy and temperature of multi-horizons black holes}

\Author{Wei Xu $^{1,2}$*, Jia Wang $^{2}$ and Xin-he Meng $^{2,3}$}

\address{%
$^{1}$ School of Physics, Huazhong University of Science and Technology, Wuhan 430074, China\\
$^{2}$ School of Physics, Nankai University, Tianjin 300071, China\\
E-mails: wangjia2010@mail.nankai.edu.cn; xhm@nankai.edu.cn\\
$^{3}$ State Key Laboratory of ITP, ITP-CAS, Beijing 100190, China}

\corres{E-mail: xuweifuture@gmail.com.}

\abstract{We present some entropy and temperature relations of multi-horizons, even including the ``virtual'' horizon. These relations are related to product, division and sum of entropy and temperature of multi-horizons. We obtain the additional thermodynamic relations of both static and rotating black holes in three and four dimensional (A)dS spacetime. Especially, a new dimensionless, charges-independence and $T_+S_+=T_-S_-$ like relation is presented. This relation does not depend on the mass, electric charge, angular momentum and cosmological constant, as it is always a constant. These relations lead us to get some interesting thermodynamic bound of entropy and temperature, including the Penrose inequality which is the first geometrical inequality of black holes. Besides, based on these new relations, one can obtain the first law of thermodynamics and Smarr relation for all horizons of black hole.}



\begin{document}


\section{Introduction}

One of central issues in quantum gravity is to understand microscopically the entropy of a black hole. Significant insights have been achieved for four and five dimensional supersymmetric, asymptotically flat, multi-charged black holes \cite{Strominger:1996sh}, where the microscopic degrees of freedom can be explained in
terms of a two-dimensional conformal field theory. Another important work has focused on the microscopic entropy of
extreme rotating solutions \cite{Guica:2008mu} via the Kerr/CFT correspondence.
Besides, it is found in \cite{Corda:2012tz,Corda:2012dw,Corda:2013gva,Corda:2013nza} that black hole entropy and the number of micro-states result to be function of the principal quantum number of the quasi-normal modes (QNMs) which can represent the ``quantum level" of black hole.
This seems consistent with various quantum gravity models where the spacetime is fundamentally discrete and leads to another possible way for studying black hole entropy microscopically.
However, the detailed
microscopic origin of the entropy of non-extremal rotating
charged black holes remains an open problem now. For this issue, much attention had been paid to the additional entropy relations of black holes recently \cite{Cvetic:2010mn,Toldo:2012ec,Cvetic:2013eda,Lu:2013ura,Chow:2013tia,
Detournay:2012ug,Castro:2012av,Visser:2012zi,Chen:2012mh,Castro:2013kea,Visser:2012wu,Abdolrahimi:2013cza,Pradhan:2013hqa,Pradhan:2013xha,
Castro:2013pqa,Faraoni:2012je,Lu:2013eoa,Anacleto:2013esa,Xu:2014qaa,Wang:2013nvz,Wang:2013smb,Xu:2013zpa,Du:2014kpa}.

Entropy product of multi-horizons \cite{Cvetic:2010mn,Toldo:2012ec,Cvetic:2013eda,Lu:2013ura,Chow:2013tia,
Detournay:2012ug,Castro:2012av,Visser:2012zi,Chen:2012mh,Castro:2013kea,Visser:2012wu,Abdolrahimi:2013cza,Pradhan:2013hqa,Pradhan:2013xha,
Castro:2013pqa,Faraoni:2012je,Lu:2013eoa,Anacleto:2013esa,Xu:2014qaa} are firstly introduced, which are expected to not only be expressed solely in terms of the quantized charges including the electric charge $Q$, the angular momentum $J$, and the cosmological constant $\Lambda$ (which can be treated as pressure after explaining the
mass of the black hole as enthalpy rather than internal energy of the system), but also have the mass independence. This study was generalize to many theories, including the super-gravity model \cite{Cvetic:2010mn,Toldo:2012ec,Cvetic:2013eda,Lu:2013ura,Chow:2013tia}, Einstein gravity \cite{Detournay:2012ug,Castro:2012av,Visser:2012zi,Chen:2012mh,Castro:2013kea,Visser:2012wu,Abdolrahimi:2013cza,Pradhan:2013hqa,Pradhan:2013xha,Xu:2014qaa} and other modified gravity models \cite{Castro:2013pqa,Cvetic:2013eda,Faraoni:2012je,Lu:2013eoa,Anacleto:2013esa,Xu:2014qaa} in both four and higher dimensions. It is always independent of the mass of the black hole as shown in \cite{Cvetic:2010mn,Castro:2012av,Toldo:2012ec,Chen:2012mh,Visser:2012zi,Cvetic:2013eda,Abdolrahimi:2013cza,Lu:2013ura,
Anacleto:2013esa,Chow:2013tia,Castro:2013kea,Lu:2013eoa,Wang:2013smb,Xu:2013zpa}. However, the mass independence of entropy product fails in some asymptotically non-flat spacetime or modified gravity theories \cite{Faraoni:2012je,Castro:2013pqa,Detournay:2012ug,Visser:2012wu,Xu:2014qaa}. Hence, the ``part'' entropy product \cite{Xu:2014qaa,Visser:2012wu,Wang:2013nvz} i.e. $\sum_{1\leq i<j\leq D}(S_{i}S_{j})^{\frac{1}{d-2}}$, here and below $D$ and $d$ are denoted as the number of horizons and the dimensions respectively in this paper, and entropy sum \cite{Xu:2014qaa,Wang:2013smb,Xu:2013zpa,Du:2014kpa} are introduced, which always are independent of the mass of the black hole in (A)dS spacetime. These two never depend on the electric charge $Q$ and angular momentum $J$, but only depend on the cosmological constant and the constants characterizing the strength of these extra matter field. Beside, it is found that entropy product and ``part'' entropy product belong to the same kind of entropy relation, as  their mass independence of entropy hold complementary in the theory without Maxwell field \cite{Xu:2014qaa}. Furthermore, ``part'' entropy product and entropy sum of Schwarzschild-de-Sitter black hole are actually equal \cite{Xu:2014qaa}, when only the effect of the physical horizons are considered, as they both can be simplified into a mass independent entropy relations of physical horizon \cite{Visser:2012wu,Xu:2014qaa}.

Actually, there exist several reasons why people study the entropy relations of multi-horizons and the physics near other horizons. Firstly, it is shown in \cite{Cvetic:1997uw,Cvetic:1997xv,Cvetic:2009jn} that the Green functions are sensitive to the geometry near all the black hole horizons, and not just the outermost one. Hence, the entropy at each horizon can therefore be expected to play a role in governing the properties of the black hole at the microscopic level. Secondly, the entropy inequalities of multi-horizons of four dimensional general axisymmetric stationary solutions in Einstein-Maxwell theory \cite{Ansorg:2010ru,Hennig:2009aa,Ansorg:2009yi,Hennig:2008zy,Ansorg:2008bv,Ansorg:2007fh,Jaramillo:2007mi} are interpreted as a general criterion for extremality \cite{Booth:2007wu}, which also results in a No-Go theorem for the possibility of force balance between two rotating black holes \cite{Hennig:2010hz}. This makes the physics at each horizon more intriguing. Besides, the effect of the other horizons is necessary in order to preserve the mass independence \cite{Visser:2012wu,Wang:2013smb,Xu:2013zpa,Du:2014kpa}. Only including this effect, these additional equalities of multi-horizons of black holes are ``universal''.

In this paper, we study the additional thermodynamic relations of black holes with multi-horizons, in order to make a further study on understanding the origin of black hole entropy at the microscopic level. The thermodynamic relations are related to the thermodynamic quantities, especially for the temperature and entropy for two physical horizons \cite{Detournay:2012ug,Chen:2012mh,Chen:2012yd,Chen:2012ps,Chen:2012pt,Chen:2013rb,Chen:2013aza,Castro:2013pqa,Chen:2013qza,
Chow:2013tia,Okamoto:1992}. Other quantities like charge potential and angular velocity are sometime considered as well \cite{Castro:2013pqa,Chow:2013tia,Pradhan:2013xha,Pradhan:2013hqa,Okamoto:1992}. These thermodynamic relations were understood well and physically, which were also linked closely with entropy product via their holographic description, i.e. the thermodynamic method of black hole/CFT correspondence \cite{Chen:2012mh,Chen:2012yd,Chen:2012ps,Chen:2012pt,Chen:2013rb,Chen:2013aza,Chen:2013qza}. The black hole/CFT (BH/CFT) correspondence focus on black holes with two physical horizons and its key point is that the inner horizon thermodynamics may play an essential role in setting up the BH/CFT correspondence. From the thermodynamics laws of both horizons (one need note that it is a so-called ``thermodynamics law''-like relation for the inner horizon), it is straightforward to read the thermodynamics of left- and right-mover and the corresponding dual temperatures. This method has only been used for the black holes in the Einstein(-Maxwell) gravity with two physical horizons up till now \cite{Chen:2012mh,Chen:2012yd,Chen:2012ps,Chen:2012pt,Chen:2013rb,Chen:2013aza,Chen:2013qza}. However, only for the two physical horizons case, one can prove that the first law of thermodynamics of the outer horizon always indicates that of the inner horizon under reasonable assumption. Moreover, only for this class of gravity theory, which is diffeomorphism invariant, its dual two-dimensional CFT is required to have $c_R = c_L$. As a result, the central charge being the same is equivalent to the condition $T_+S_+=T_-S_-$ in the thermodynamics method, or equivalently the condition that the entropy product $S_+S_-$ being mass-independent \cite{Detournay:2012ug,Chen:2012mh,Chen:2012yd,Chen:2012ps,Chen:2012pt,Chen:2013rb,Chen:2013aza,Chen:2013qza}, where $T_{\pm}$, $S_{\pm}$ are the outer and inner horizon temperatures and entropies respectively. Therefore the thermodynamics relations $T_+S_+=T_-S_-$ (equivalently entropy product $S_+S_-$ being mass-independent) may be taken as the criterion whether there is a 2 dimensional CFT dual for the black holes in the Einstein gravity and other diffeomorphism invariant gravity theories \cite{Chen:2012mh,Chen:2012yd,Chen:2012ps,Chen:2012pt,Chen:2013rb,Chen:2013aza,Chen:2013qza}. Namely, these thermodynamics relations really give a clue to reveal the microscopics of black holes. When the discussion generalizes to some cases of black holes with more than two horizons, the thermodynamics relations $T_+S_+=T_-S_-$ (equivalently entropy product $S_+S_-$ being mass-independent) breaks down \cite{Castro:2013pqa,Detournay:2012ug,Visser:2012wu,Xu:2014qaa,Chen:2012mh}, even there are only two physical horizons. Hence, we aim to study the additional thermodynamic relations of black holes with more than two horizons (other than the case of two physical horizons), in order to improve the study on understanding the origin of black holes entropy at the microscopic level for these (A)dS black holes.

Mainly, we present some general entropy and temperature relations of multi-horizons, even the ``virtual'' horizon. Actually, studying this kind of relations for understanding the inside physics of black holes is not an new idea. The discussion for Kerr black hole had been presented in \cite{Okamoto:1992} in 1992. These relations are related to product, division and sum of entropy and temperature of multi-horizons. We generalize them to the similar relations of static and rotating black holes in three and four dimension. Some general thermodynamic relations are constructed and found to be hold for both AdS and dS black holes. Moreover, a new dimensionless and charge-independence relation $\left(\sum_{i=1}^D T_iS_i\right)\times\left(\sum_{i=1}^D\frac{1}{T_iS_i}\right)$
is presented, which is the generalizing of the $T_+S_+=T_-S_-$ relations for two horizons. It does not depend on the mass, electric charge, angular momentum and cosmological constant, as it is always a constant for black holes as shown in the present papers. This relation is expected to be helpful of constructing BH/CFT correspondence for more-than-two-horizons, in order to understand microscopically the black hole entropy. Besides, based on these relations, we obtain some interesting thermodynamic bound of entropy and temperature, including the Penrose inequality which is the first geometrical inequality of black hole (event horizon) (See \cite{Mars:2009cj} as a review article). We also get the Penrose-like inequalities for other horizons. Moreover, these thermodynamic relations can lead to the thermodynamic law and Smarr relation of event horizon and Cauchy horizon of black holes. We take the case of Kerr black hole as a detailed example. This discussion is generalized to Kerr-Newman black hole as well.

This paper is organized as follows. In the next Section, we will investigate the thermodynamic relations of static black holes in four dimensional (A)dS spacetime. In Section 3, the thermodynamic relations of rotating black holes in three dimensions and four dimensions are presented. Using these thermodynamic relations, we obtain some interesting thermodynamic bound in Section 4 and the first law and Smarr relation of black holes in Section 5, respectively. Section 6 is devoted to the conclusions and discussions.


\section{Thermodynamic relations of static (A)dS black holes}

In the section, we investigate many thermodynamic relations of static black holes in four dimensional (A)dS spacetime. Among these relations, a new dimensionless and charge-independence relation $\left(\sum_{i=1}^D T_iS_i\right)\times\left(\sum_{i=1}^D\frac{1}{T_iS_i}\right)$ is presented, which is the generalizing of the $T_+S_+=T_-S_-$ relations.

\subsection{Thermodynamic relations of Schwarzschild-(A)dS black holes}\label{section2.1}

We firstly give the thermodynamic relations of Schwarzschild-(A)dS black hole in detail. We begin with the line element
\begin{align*}
ds^2=-f(r)dt^2+\frac{dr^2}{f(r)}+r^{2}\left(d\theta^{2}+\sin^{2}d\varphi^{2}\right).
\end{align*}
For four dimensional Schwarzschild-de-Sitter black hole, the horizon function is
\begin{align*}
f(r)=1-\frac{2M}{r}-\frac{\Lambda r^2}{3},
\end{align*}
with $M$ being the mass of the black hole and $\Lambda=\frac{1}{L^2}$ being the cosmological constant. The three black hole horizons are \cite{Visser:2012wu}
\begin{align}
r_{E} &=2L\sin\left(\frac 13 \arcsin \left(\frac{3 M}{L}\right)\right)\nonumber\\
r_{C} &=2L\sin\left(\frac 13 \arcsin \left(\frac{3 M}{L}\right)+\frac{2\pi}{3}\right)\nonumber\\
r_{V} &=2L\sin\left(\frac 13 \arcsin \left(\frac{3 M}{L}\right)-\frac{2\pi}{3}\right),\label{sch-horizon}
\end{align}
where, $r_{E}$, $r_{C}$ and $r_V$ represent the event horizon, cosmological horizon and ``virtual'' horizon respectively. The Bekenstein area-entropy \cite{Bekenstein:1972tm} and Hawking temperature \cite{Hawking:1974sw} of each horizon are $S_{i}=A_{i}/4=\pi r_i^2$ and
\begin{align*}
T_i=\frac{f^{\prime}(r_i)}{4\pi}=\frac{L^2-r_i^2}{4\pi L^2r_i},\ (i=E,C,V)
\end{align*}
respectively, where $f^{\prime}(r)$ denotes the derivative function of $f(r)$ respect to $r$.

We firstly revisit some known entropy relations: the mass-dependence entropy product \cite{Visser:2012wu}
\begin{align}
  S_ES_CS_V=36\pi^3M^2L^4;\label{sch-product}
\end{align}
the equal and mass-independence ``part'' entropy product and entropy sum \cite{Visser:2012wu,Wang:2013smb,Xu:2014qaa}
\begin{align}
  &S_ES_C+S_ES_V+S_CS_V=9\pi^2L^4;\label{sch-part}\\
  &S_E+S_C+S_V=6 \pi L^2\label{sch-sum},
\end{align}
which can lead into the entropy relations of two physical horizon having mass independence \cite{Visser:2012wu,Xu:2014qaa}
\begin{align}
  S_E+S_C+\sqrt{S_ES_C}=3\pi L^2.\label{sch-equal}
\end{align}
Basing on these entropy relations, one can construct and calculate more relations. For example, the mass-independence case:
\begin{align}
  S_E^2+S_C^2+S_V^2&=(S_E+S_C+S_V)^2-2(S_ES_C+S_ES_V+S_CS_V)\nonumber\\
  &=18\pi^2L^4;
\end{align}
and some mass-dependence cases
\begin{align}
&\frac{1}{S_E}+\frac{1}{S_C}+\frac{1}{S_V}=\frac{1}{4\pi M^2},\label{sch-mass1}\\
&\frac{1}{S_ES_C}+\frac{1}{S_ES_V}+\frac{1}{S_CS_V}=\frac{1}{6\pi^2M^2L^2}, \nonumber\\
&\frac{S_CS_V}{S_E}+\frac{S_VS_E}{S_C}+\frac{S_ES_C}{S_V}=\frac{9\pi L^4}{4M^2}-12\pi L^2, \nonumber\\
&\frac{S_C+S_V}{S_E}+\frac{S_V+S_E}{S_C}+\frac{S_E+S_C}{S_V}=\frac{3L^2}{2M^2}-3,\nonumber
\end{align}
following the same procedure.

Then we turn to the temperature relations and study the product, division and sum of temperature of multi-horizons. In order to construct similar relations, we need to introduce the relationship of three horizons
\begin{align*}
  &r_E+r_C+r_V=0,\quad r_E+r_C+r_V=-6ML^2\\
  &r_Er_C+r_Er_V+r_Cr_V=-3L^2,
\end{align*}
by using which, one can find the mass-independence case:
\begin{align*}
  &T_ET_C+T_CT_V+T_VT_E=0, \\
  &\frac{1}{T_E}+\frac{1}{T_C}+\frac{1}{T_V}=0, \\
  &\frac{T_C+T_V}{T_E}+\frac{T_V+T_E}{T_C}+\frac{T_E+T_C}{T_V}=-3,
\end{align*}
and some mass-dependence cases
\begin{align}
&T_E+T_C+T_V=\frac{1}{8\pi M}, \nonumber\\
&T_E^2+T_C^2+T_V^2=\frac{1}{64\pi^2M^2},\label{sch-tem} \\
&\frac{T_CT_V}{T_E}+\frac{T_VT_E}{T_C}+\frac{T_ET_C}{T_V}=-\frac{1}{4\pi M},\nonumber \\
&\frac{1}{T_ET_C}+\frac{1}{T_CT_V}+\frac{1}{T_VT_E}=\frac{12\pi^2L^4}{9M^2-L^2},\nonumber\\
&T_ET_CT_V=\frac{6M}{L^4}-\frac{2}{3ML^2}.\nonumber
\end{align}
One can conclude that
\begin{align*}
  &\frac{1}{S_E}+\frac{1}{S_C}+\frac{1}{S_V}=16\pi(T_E+T_C+T_V)^2=16\pi(T_E^2+T_C^2+T_V^2)
\end{align*}

For the $T_+S_+=T_-S_-$ like relations, one can find:
\begin{align}
&T_ES_E+T_CS_C+T_VS_V=\frac{9M}{2}, \\
&\frac{1}{T_ES_E}+\frac{1}{T_CS_C}+\frac{1}{T_VS_V}=\frac{2}{M},
\end{align}
which are both dependent of mass and different from the results for flat black holes with two physical horizons.

For the discussion of Schwarzschild-AdS black hole, one can find the above relations are all universal after taking the transitions $L\rightarrow iL$. Namely, instead $L$ of $\frac{1}{\Lambda}$, all the above relations always hold for Schwarzschild-(A)dS black hole. Moreover, we obtain the dimensionless relation
\begin{align}
&(T_ES_E+T_CS_C+T_VS_V)\left(\frac{1}{T_ES_E}+\frac{1}{T_CS_C}+\frac{1}{T_VS_V}\right)=9.
\end{align}
One will find that this one is always charge-independence, as shown again in the cases for Reissner-Nordstr\"om-(A)dS black holes.

\subsection{Thermodynamic relations of Reissner-Nordstr\"om-(A)dS black holes}\label{section2.2}

For a further studying, we list the thermodynamic relations of Reissner-Nordstr\"om-(A)dS black holes following the same procedure. The horizon function is
\begin{align*}
f(r)=1-\frac{2M}{r}-\frac{\Lambda r^2}{3}+\frac{Q^2}{r^2},
\end{align*}
The four black hole horizons are $r_1$, $r_2$, $r_3$ and $r_4$. The Bekenstein area-entropy \cite{Bekenstein:1972tm} and Hawking temperature \cite{Hawking:1974sw} of each horizon are $S_{i}=A_{i}/4=\pi r_i^2$ and
\begin{align*}
T_i=\frac{f^{\prime}(r_i)}{4\pi}=\frac{1}{4\pi}\left(\frac{1}{r_i}-\Lambda r_i-\frac{Q^2}{r_i^3}\right),\ (i=1,2,3,4)
\end{align*}
respectively.

We still firstly revisit some known entropy relations: the mass-independence entropy product and entropy sum\cite{Visser:2012wu,Wang:2013smb}
\begin{align}
  &S_1S_2S_3S_4=\frac{9\pi^4Q^4}{\Lambda^2}\\
  &S_1+S_2+S_3+S_4=\frac{6 \pi}{\Lambda}.
\end{align}
Basing on these entropy relations and the following relationship of four horizons
\begin{align*}
  &r_1+r_2+r_3+r_4=0,\quad r_1r_2+r_1r_3+r_1r_4+r_2r_3+r_2r_4+r_3r_4=-\frac{3}{\Lambda},\\
  &r_1r_2r_3r_4=-\frac{3Q^2}{\Lambda},\quad r_1r_2r_3+r_1r_3r_4+r_1r_2r_4+r_2r_3r_4=-\frac{6M}{\Lambda},
\end{align*}
one can also construct and calculate more relations. For example, the mass-independence case:
\begin{align*}
 &S_1S_2+S_1S_3+S_2S_3+S_1S_4+S_2S_4+S_3S_4=\frac{9\pi^2}{\Lambda^2}-\frac{6\pi^2Q^2}{\Lambda},\\
 &\frac{1}{S_1S_2}+\frac{1}{S_1S_3}+\frac{1}{S_2S_3}+\frac{1}{S_1S_4}+\frac{1}{S_3S_4}+\frac{1}{S_2S_4}=\frac{1}{\pi^2Q^4}-\frac{2\Lambda}{3\pi^2Q^2};
\end{align*}
and some mass-dependence cases
\begin{align*}
&S_1S_2S_3+S_1S_3S_4+S_2S_3S_4+S_1S_2S_4=-\frac{6\pi^3(M^2+3Q^2)}{\Lambda^2},\\
&S_1^2+S_2^2+S_3^2+S_4^2=\frac{18\pi^2}{\Lambda^2}+\frac{12\pi^2Q^2}{\Lambda},\\
&\frac{1}{S_1}+\frac{1}{S_2}+\frac{1}{S_3}+\frac{1}{S_4}=-\frac{2(M^2+3Q^2)}{3\pi Q^4},\\
&\frac{S_2S_3S_4}{S_1}+\frac{S_1S_3S_4}{S_2}+\frac{S_1S_2S_4}{S_3}+\frac{S_2S_3S_4}{S_4}=\,{\frac {4\pi^2{M}^{4}}{{\Lambda}^{2}{Q}^{4}}}+\,{\frac {24\pi^2{M}^{2}}{{\Lambda}^{2}{Q}^{2}}}+\,{\frac {12\pi^2{Q}^{2}}{\Lambda}}+\frac{18\pi^2}{\Lambda^2}, \\
&\frac{S_2+S_3+S_4}{S_1}+\frac{S_1+S_3+S_4}{S_2}+\frac{S_1+S_2+S_4}{S_3}+\frac{S_1+S_2+S_3}{S_4}=-\,{\frac {4{M}^{2}}{\Lambda\,{Q}^{4}}}-\,{\frac {12}{\Lambda\,{Q}^{2}}}-4,
\end{align*}
following the same procedure.

Then we turn to the temperature relations and study the product, division and sum of temperature of multi-horizons. However, one can only find some mass-dependence cases:
\begin{align*}
&T_1+T_2+T_3+T_4=\,{\frac {2M}{\pi{Q}^{2}}}-{\frac {13{M}^{3}}{3\pi{Q}^{4}}}, \\
&T_1T_2T_3+T_1T_3T_4+T_2T_3T_4+T_1T_2T_4=\,{\frac {7\Lambda\,{M}^{3}}{36\pi^3{Q}^{4}}}.
\end{align*}
These relations are not universal as one anticipate.

For the $T_+S_+=T_-S_-$ like relations, one can find:
\begin{align}
&T_1S_1+T_2S_2+T_3S_3+T_4T_4=16M, \\
&\frac{1}{T_1S_1}+\frac{1}{T_2S_2}+\frac{1}{T_3S_3}+\frac{1}{T_4S_4}=0,
\end{align}
which are both from the results for asymptotically flat black holes with two physical horizons and that for Schwarzschild-(A)dS black holes.

One can note that all the above relations always hold for both AdS and dS black holes. Moreover, we again obtain the dimensionless and charge-independence relation
\begin{align} \label{eq:RN_dimless}
(T_1S_1+T_2S_2+T_3S_3+T_4S_4)\left(\frac{1}{T_1S_1}+\frac{1}{T_2S_2}+\frac{1}{T_3S_3}+\frac{1}{T_4S_4}\right)=0,
\end{align}
for Reissner-Nordstr\"om-(A)dS black holes. This also belongs to the kind of thermodynamic relation $\left(\sum_{i=1}^D T_iS_i\right)\times\left(\sum_{i=1}^D\frac{1}{T_iS_i}\right)$
with $D=4$ being the number of the horizons, which is always a constant. Consider the degenerated case $\Lambda=0$, i.e. Reissner-Nordstr\"om black hole, the relation (\ref{eq:RN_dimless}) reduce to
\begin{align}
(T_1S_1+T_2S_2)\left(\frac{1}{T_1S_1}+\frac{1}{T_2S_2}\right)=0,
\end{align}
$D=2$ in this case, one is event horizon and another is Cauchy horizon, is also a constant.
This is the generalizing of the $T_+S_+=T_-S_-$ relations, hence it is expected to be helpful of constructing BH/CFT correspondence for more than two horizons.

\section{Thermodynamic relation of rotating black holes}

In this section, we focus on thermodynamic relation of rotating black holes in three and four dimensions, especially for the the dimensionless relation. Namely, we will mainly recheck the generalizing of the $T_+S_+=T_-S_-$ relations for more than two horizons, i.e. $\left(\sum_{i=1}^D T_iS_i\right)\times\left(\sum_{i=1}^D\frac{1}{T_iS_i}\right)$
for rotating black holes. It is still a constant, which is charge-independent (mass-independent, charge-independent and angular-momentum-independent).

\subsection{Thermodynamic relation of Kerr(-Newman) black hole}\label{section3.1}

The Kerr metric in Boyer-Linquist coordinates is
\begin{align*}
ds^2=&-\frac{\Delta-a^2\sin^2\theta}{\Sigma}dt^2-2a\sin^2\theta\left(\frac{r^2+a^2-\Delta}{\Sigma}\right)dtd\phi\notag\\
&+\left(\frac{(r^2+a^2)^2-\Delta a^2\sin^2\theta}{\Sigma}\right)\sin^2\theta d^2\phi+\frac{\Sigma}{\Delta}dr^2
 +\Sigma d^2\theta,
\end{align*}
where
\begin{align*}
\Sigma &=r^2+a^2\cos^2\theta, \\
\Delta(r) &=r^2-2 M r+a^2.
\end{align*}
The zeros of $\Delta(r)$ correspond to the event horizon and Cauchy horizon, denoted as $r_{+}$ and $r_{-}$
\begin{align}
r_{+}=M+\sqrt{M^2-a^2},\nonumber \\
r_{-}=M-\sqrt{M^2-a^2}.\label{kerr-horizon}
\end{align}
Since the metric is not diagonal, the Hawking temperature \cite{Hawking:1974sw} should be $T=\frac{\kappa}{2\pi}$. The surface gravity
$\kappa$ and temperature $T$ are
\begin{align}
&\kappa_{\pm}=\frac{r_{\pm}-r_{\mp}}{2(r_{\pm}^2+a^2)}, \notag \\
&T_{\pm}=\frac{r_{\pm}-r_{\mp}}{4\pi(r_{\pm}^2+a^2)}.
\end{align}
And the Bekenstein area-entropy \cite{Bekenstein:1972tm} for each horizon is
\begin{align}
S_{\pm}=\frac{A_{\pm}}{4}=\pi(r_{\pm}^2+a^2),
\label{Kerrarea}
\end{align}
with
\begin{align}
  T_{\pm}S_{\pm}=\frac{r_{\pm}-r_{\mp}}{4}.
\end{align}
Obviously, these lead to the entropy product
\begin{align}
S_{+}S_{-}=4\pi^2J^2,
\label{Kerrproduct}
\end{align}
entropy sum
\begin{align}
S_{+}+S_{-}=4\pi\,M^2,
\label{Kerrsum}
\end{align}
entropy minus
\begin{align}
S_{+}-S_{-}=2\pi\,M(r_{+}-r_{-})=\pm8\pi\,MT_{\pm}S_{\pm},
\label{Kerrminus}
\end{align}
and the mass-dependent relation
\begin{align}
\frac{1}{S_{+}}+\frac{1}{S_{-}}=\frac{S_{+}+S_{-}}{S_{+}S_{-}}=\frac{\,M^2}{\pi\,J^2},
\label{Kerrversum}
\end{align}
where $J=aM$ is the angular momentum.
The angular velocity is
\begin{align}
\Omega_{\pm}=\frac{a}{r_{\pm}^2+a^2}=\frac{\pi\,J}{MS_{\pm}},
\label{KerrOmega}
\end{align}
based on which, we can obtain
\begin{align}
\Omega_{+}+\Omega_{-}=\frac{M}{J}
\label{KerrOmegasum}
\end{align}
On the other hand, we find
\begin{align}
T_{+}S_{+}+T_{-}S_{-}=0,\label{kerrTS} \\ \frac{1}{T_{+}S_{+}}+\frac{1}{T_{-}S_{-}}=0\nonumber,
\end{align}
thus
\begin{align} \label{eq:Kerr_dimless}
(T_{+}S_{+}+T_{-}S_{-})\left(\frac{1}{T_{+}S_{+}}+\frac{1}{T_{-}S_{-}}\right)=0.
\end{align}
which shows that the charge-independent of dimensionless relation $\left(\sum_{i=1}^D T_iS_i\right)\times\left(\sum_{i=1}^D\frac{1}{T_iS_i}\right)$ holds.

For Kerr-Newman black hole, one can substitute $a^2$ with $a^2+Q^2$, where $Q$ is the electric charge, apparently the relation (\ref{eq:Kerr_dimless}) still holds.

\subsection{Thermodynamic relation of BTZ Black Hole}\label{section3.2}

Considering the BTZ black hole \cite{Banados:1992wn}, the metric reads
\begin{align*}
ds^2=-f(r)dt^2+\frac{dr^2}{f(r)}+r^2(N^{\phi}(r)dt+d\phi)^2,
\end{align*}
where the cosmological constant $\Lambda=-\frac{1}{l^2}$, and the horizon function $f(r)$ and the angular velocity $N^{\phi}(r)$
\begin{align*}
f(r)&=-M+\frac{r^2}{\ell^2}+\frac{J^2}{4r^2}, \\
N^{\phi}(r)&=-\frac{J}{2r^2},
\end{align*}
where $M$ and $J$ are the mass and angular momentum of the black hole respectively. The horizon function $f(r)$
has four zeros
\begin{align*}
r_1&=\sqrt{\left(1+\sqrt{1-\left(\frac{J}{M\ell}\right)^2}\right)M\ell},\quad r_2=-\sqrt{\left(1+\sqrt{1-\left(\frac{J}{M\ell}\right)^2}\right)M\ell},\\
r_3&=\sqrt{\left(1-\sqrt{1-\left(\frac{J}{M\ell}\right)^2}\right)M\ell},\quad r_4=-\sqrt{\left(1-\sqrt{1-\left(\frac{J}{M\ell}\right)^2}\right)M\ell},
\end{align*}
where $r_1$ and $r_3$ correspond to event horizon and Cauchy horizon, i.e. physical horizons, while $r_2$ and
$r_4$ represent the negative and un-physical, ``virtual'' horizons which often is discarded in literature.

As discussed in \cite{Banados:1992wn}, the Bekenstein ``area''-entropy \cite{Bekenstein:1972tm} is equal to twice the perimeter length of the horizon,
\begin{align*}
S_i=4\pi r_i
\end{align*}
and Hawking temperature \cite{Hawking:1974sw} is
\begin{align*}
T_i=\left(\frac{\partial S_i}{\partial M}\right)^{-1}.
\end{align*}
Firstly, we discuss the physical horizons only, i.e. $r_1$ and $r_3$. With the help of computer algebra system
(CAS) one can get
\begin{align}
T_1S_1+T_3S_3=0,
\end{align}
meanwhile $T_1S_1$ or $T_3S_3$ does not equal to zero explicitly, so obviously
\begin{align}
(T_1S_1+T_3S_3)\left(\frac{1}{T_1S_1}+\frac{1}{T_3S_3}\right)=0.
\end{align}
Secondly, we should include the un-physical, ``virtual'' horizons $r_2$ and $r_4$. We note that $r_1$ and $r_2$
have a opposite sign as well as $r_3$ and $r_4$. So $S_2$, $S_4$, $T_2$ and $T_4$ have a minus sign to their
counterpart respectively we discussed in last paragraph. However, $T_2S_2+T_4S_4=T_1S_1+T_3S_3=0$, apparently
\begin{align}
(T_1S_1+T_2S_2+T_3S_3+T_4S_4)\left(\frac{1}{T_1S_1}+\frac{1}{T_2S_2}+\frac{1}{T_3S_3}+\frac{1}{T_4S_4}\right)=0.
\end{align}
Again the dimensionless relation $\left(\sum_{i=1}^DT_iS_i\right)\times\left(\sum_{i=1}^D\frac{1}{T_iS_i}\right)$ is charge-independent.

\section{Thermodynamic relations and thermodynamic bound}

In this section, based on the thermodynamic relations presented in this paper, we obtain some thermodynamic bounds of the thermodynamic quantities including the entropy and temperature. From the entropy bound, one can get the area bound. Especially for the upper area bound of the event horizon of black holes, one can find that it is actually the exact Penrose inequality of black hole, which is the first geometrical inequality of black hole (See \cite{Mars:2009cj} as a review article).

\subsection{Thermodynamic bound for Schwarzschild-dS black hole}

Firstly we show the thermodynamic bound for Schwarzschild-dS black hole. Consider the thermodynamic relations of Schwarzschild-dS black hole shown in Section \ref{section2.1}. We will only focus on the cases with \begin{align}
\frac{3M}{L}\leq1,\label{condition}
\end{align}
which leads to three real horizons according to Eq.(\ref{sch-horizon}). Then we will find the entropy and temperature for three horizons are all real. (For $\frac{3M}{L}>1$, one will only find one real horizon.) Especially for the event horizon, cosmological horizon and the ``virtual" horizon, we know that $0\leq\,r_E\leq\,L\leq\,r_C\leq|r_V|\leq2L$ and $r_V<0$, hence $0\leq\,S_E\leq\,S_C\leq\,S_V\leq4\pi\,L^2$, and $T_E\geq0, T_C\leq0, T_V\geq0$ (Note the Hawking temperature of cosmological horizon is $\hat{T}_C=-T_C\geq0$).

From thermodynamic relation Eq.(\ref{sch-equal}), we get
\begin{align*}
  0\leq3S_E\leq(S_E+S_C+\sqrt{S_ES_C})=3\pi\,L^2\leq3\,S_C,
\end{align*}
and
\begin{align*}
  0\leq\,S_C\leq3\pi\,L^2,
\end{align*}
which together give
\begin{align*}
  0\leq\,S_E\leq\pi\,L^2\leq\,S_C\leq3\pi\,L^2.
\end{align*}
As $0\leq\,(S_C+S_E)\leq3\pi\,L^2$, from the entropy sum Eq.(\ref{sch-sum}), we find
\begin{align*}
  \,S_V\geq3\pi\,L^2.
\end{align*}

Totaly, we obtain the entropy bound of the event horizon, the cosmological horizon and the negative horizon
\begin{align}
\,S_E\in\bigg[0,\pi\,L^2\bigg],\quad\,S_C\in\bigg[\pi\,L^2,3\pi\,L^2\bigg],\quad\,S_V\in\bigg[3\pi\,L^2,4\pi\,L^2\bigg].
\end{align}
And the area entropy leads to the area bound
\begin{align}
\sqrt{\frac{\,A_E}{16\pi}}\in\bigg[0,\frac{\,L}{2}\bigg],\quad\,\sqrt{\frac{\,A_C}{16\pi}}\in\bigg[\frac{L}{2},\sqrt{\frac{3}{4}}\,L\bigg],\quad\,\sqrt{\frac{\,A_V}{16\pi}}\in\bigg[\sqrt{\frac{3}{4}}\,L,L\bigg],
\end{align}
which are all geometrical bounds of black hole horizons, as parameter $L$ is actually the cosmological radius.

On the other hand, consider the mass-dependence thermodynamic relations of entropy Eq.(\ref{sch-mass1}), we can also find
\begin{align*}
  S_E\geq4\pi\,M^2,
\end{align*}
which leads to the Penrose-like inequality
\begin{align}
  \sqrt{\frac{A_E}{16\pi}}\geq\frac{M}{4}
\end{align}
Inserting condition Eq.(\ref{condition}) into the area bound, we can obtain other Penrose-like inequalities
\begin{align}
\sqrt{\frac{\,A_C}{16\pi}}\geq\frac{3M}{2},\quad\,\sqrt{\frac{\,A_V}{16\pi}}\geq3\sqrt{\frac{3}{4}}\,M.
\end{align}
Furthermore, we can get the temperature bound from Eq.(\ref{sch-tem}), i.e.
\begin{align}
T_i\leq\frac{1}{8\pi\,M}\,(i=E,V);\quad\,|T_C|\leq\frac{1}{8\pi\,M}
\end{align}

\subsection{Thermodynamic bound for Kerr black holes}

For black holes in rotating spacetime, we take the thermodynamic relations of the Kerr black holes as an example, which is shown in Section \ref{section3.1}. From Eq.(\ref{kerr-horizon}), the existence of black hole horizons leads to the famous Kerr bound
\begin{align}
\frac{M}{a}\geq\,1,
\end{align}
or equivalently $M^2\geq\,J$.
Besides, $r_+\geq\,r_-$ leads to $S_+\geq\,S_-\geq\,0$. Then the entropy product Eq.(\ref{Kerrproduct}) results in
\begin{align*}
S_+\geq\sqrt{S_+S_-}=2\pi\,J,\quad\,S_-\leq\sqrt{S_+S_-}=2\pi\,J.
\end{align*}
And from the entropy sum Eq.(\ref{Kerrsum}), we obtain
\begin{align*}
2\pi\,M^2=\frac{(S_{+}+S_-)}{2}\leq\,S_+\leq\,(S_{+}+S_-)=4\pi\,M^2,\quad\,S_-\leq\frac{(S_{+}+S_-)}{2}=2\pi\,M^2.
\end{align*}

These together give the entropy bound of the event horizon and the Cauchy horizon
\begin{align}
S_+\in\bigg[2\pi\,M^2,4\pi\,M^2\bigg],\quad\,S_-\in\bigg[0,2\pi\,J\bigg].
\end{align}
Note the Kerr bound relation is used here.
Because of the area entropy Eq.(\ref{Kerrarea}), we also get the area bound of the event horizon and the Cauchy horizon
\begin{align}
\sqrt{\frac{A_+}{16\pi}}\in\bigg[\frac{M}{\sqrt{2}},M\bigg],\quad\,\sqrt{\frac{A_-}{16\pi}}\in\bigg[0,\sqrt{\frac{J}{2}}\bigg].
\end{align}
Note the upper bound of event horizon is actually the exact Penrose inequality of black hole.

Actually, for black holes with real horizons, (e.g. Kerr-Newman black hole, Reissner-Nordstr\"om-dS black hole, et al), we can follow the similar procedure to obtain the bound of thermodynamic quantities. However, if the black hole horizon is not real, the method presented here fails.

\section{Thermodynamic relations and thermodynamic laws}

The thermodynamic bounds in the preceding section can be seen as a application of the thermodynamic relations, while how they link to the thermodynamic of black holes seems to be still unclear. However, based on these thermodynamic relations as well, one can obtain the thermodynamic law and Smarr relation of all horizons of black holes.
In this section, we only take the discussion about the event horizon and Cauchy horizon of Kerr black hole as a detailed example.

From the entropy product Eq.(\ref{Kerrproduct}) and entropy sum Eq.(\ref{Kerrsum}), one can find
\begin{align*}
  S_{-}dS_{+}+S_{+}dS_{-}&=8\pi^2J dJ,\\
  dS_{+}+dS_{-}&=8\pi\,M dM.
\end{align*}
This leads to
\begin{align*}
  &dS_{+}=-\frac{8\pi^2J}{S_{+}-S_{-}}dJ+\frac{8\pi\,M\,S_{+}}{S_{+}-S_{-}}dM\\
  &dS_{-}=\frac{8\pi^2J}{S_{+}-S_{-}}dJ-\frac{8\pi\,M\,S_{-}}{S_{+}-S_{-}}dM.
\end{align*}
Then using the entropy minus relation Eq.(\ref{Kerrminus}), this can be transformed to
\begin{align*}
  &dM=+\,T_{+}d S_{+}+\frac{\pi\,J}{MS_{+}}dJ\\
  &dM=-\,\hat{T}_{-}d S_{-}+\frac{\pi\,J}{MS_{-}}dJ,
\end{align*}
where we have used the relation of the Hawking temperature of the Cauchy horizon \cite{Chen:2012mh}
\begin{align}
\,\hat{T}_{-}=-T_+|_{r_{+}\leftrightarrow\,r_{-}}=-T_{-}
\label{kerrCT}
\end{align}
and $r_{+}\leftrightarrow\,r_{-}$ is the exchange of two horizons.
After inserting the relations Eq.(\ref{KerrOmega}), we obtain first law of thermodynamic of the event horizon and the Cauchy horizon of Kerr black hole from the above equations
\begin{align}
  &d M=+\,T_{+}d S_{+}+\Omega_{+} d J\nonumber\\
  &d M=-\,\hat{T}_{-}d S_{-}+\Omega_{-} dJ.\label{kerrfirst}
\end{align}

On the other hand, the scaling discussion of mass $M(S_+,J)$ gives
\begin{align*}
M=2(+\,T_{+} S_{+}+\Omega_{+}J).
\end{align*}
For the Smarr relation of the Cauchy horizon, one can assuming that
\begin{align*}
M=c\,\hat{T}_{-} S_{-}+d\Omega_{-}J.
\end{align*}
Using the Hawking temperature of Cauchy horizon Eq.(\ref{kerrCT}) and the dimensionless relation Eq.(\ref{kerrTS}), the Smarr relation of the Cauchy horizon leads to \begin{align*}
2M&=2(+\,T_{+} S_{+}+\Omega_{+}J)+(-c\,T_{-} S_{-}+d\Omega_{-}J)\\
&=-(2+c)\,T_{-} S_{-}+(2\Omega_{+}+d\Omega_{-})J.
\end{align*}
Inserting the relation Eq.(\ref{KerrOmegasum}), we find
\begin{align*}
2M&=-(2+c)\,T_{-} S_{-}+\bigg((d-2)\Omega_{-}J+2M\bigg)
\end{align*}
implying $c=-2,d=2$, as the mass behaviors as $M(S_{-},\Omega_{-})$ in the Cauchy horizon.
Finally, we get the Smarr relation of the event horizon and the Cauchy horizon of Kerr black hole
\begin{align}
M=2\bigg(+\,T_{+} S_{+}+\Omega_{+}  J\bigg)\nonumber\\
M=2\bigg(-\,\hat{T}_{-} S_{-}+\Omega_{-}  J\bigg).\label{kerrSmarr}
\end{align}
The above first law of thermodynamic Eq.(\ref{kerrfirst}) and Smarr relation Eq.(\ref{kerrSmarr}) of the event horizon and the Cauchy horizon of Kerr black hole are consistent with that in \cite{Castro:2013pqa}.

This discussion can be easily generalized to Kerr-Newman black hole by substituting $a^2$ with $a^2+Q^2$ in the whole procedure. One can get the following thermodynamic laws
\begin{align}
  d M&=\pm\,\hat{T}_{\pm}d S_{\pm}+\Omega_{\pm} d J+\Phi_{\pm} d Q,\\
  M&=2\bigg(\pm\,\hat{T}_{\pm} S_{\pm}+\Omega_{\pm}  J\bigg)+\Phi_{\pm} Q.
\end{align}
Note here the Hawking temperature of Cauchy horizon Eq.(\ref{kerrCT}) is used as well.
One may expect to generalized this discussion to black holes with three horizons as well. For example, one can try to obtain the thermodynamic laws of the Schwarzschild-(A)dS black holes by these relations. However, it is not such easy to work out it, as the first law of thermodynamic and Smarr relation in (A)dS spacetime are still open questions.

\section{Conclusions}

In this paper, we study the additional thermodynamic relations of black holes with multi-horizons, in order to make a further study on understanding the origin of black hole entropy at the microscopic level. We obtain some general entropy and temperature relations of multi-horizons, even the ``virtual'' horizon. We also present how we get these relations in detailed. These relations are related to product, division and sum of entropy and temperature of multi-horizons. We consider the static and rotating black holes in three and four dimensions. Some general thermodynamic relations are constructed and found to be hold for both AdS and dS black holes. Moreover, a new dimensionless and charge-independence relation $\left(\sum_{i=1}^D T_iS_i\right)\times\left(\sum_{i=1}^D\frac{1}{T_iS_i}\right)$ is presented. It does not depend on the mass, electric charge, angular momentum and cosmological constant, because it is a constant for both static and rotating black holes. As it is the generalizing of the $T_+S_+=T_-S_-$ relations, this relation is expected to be helpful of constructing black hole/CFT (BH/CFT) correspondence for more-than-two-horizons. Besides, this dimensionless relation is invariant in an inverse transformation $TS\rightarrow \frac{1}{TS}$. This is a symmetry even if the underlying physical picture is not clear.

On the other hand, the improvement of the topic of thermodynamic relations is a little, while some attempts are shown in this work. Based on the thermodynamic relations presented in this paper, we obtain some thermodynamic bounds of the thermodynamic quantities including the entropy and temperature. Especially for the upper area bound of the event horizon of black holes, one can find that it is actually the exact Penrose inequality of black hole, which is the first geometrical inequality of black hole. We also get Penrose-like inequality of the other horizons. The thermodynamic bounds can be seen as a application of the thermodynamic relations, while how they link to the thermodynamic of black holes seems to be still unclear. However, based on these thermodynamic relations as well, we present the thermodynamic law and Smarr relation of event horizon and Cauchy horizon of Kerr black hole as a detailed example. This discussion is generalized to Kerr-Newman black hole as well. These applications of thermodynamic relations also indicate that thermodynamics of the inner horizon is linked closely with that of event horizon. This is consistent with the black hole/CFT correspondence \cite{Chen:2012mh}.

It is also interesting to generalize this discussion about the close relationships between thermodynamics of multi-horizons to that of three and higher dimensional black holes and other black holes with multi-horizons, including the thermodynamic bound and the thermodynamic laws. Especially for black holes in (A)dS spacetime (e.g. Schwarzschild-(A)dS black holes), for which the first law of thermodynamic and Smarr relation are still open questions. An interesting idea is treating the cosmological constant as a dynamic viable (see, e.g. \cite{Kastor:2009wy,Dolan:2011xt,Cvetic:2010jb,Dolan:2013ft,Xu:2013zea,Xu:2014tja,Xu:2014kwa,Altamirano:2014tva}). This may be possibly checked in this way.
These are all left to be the future tasks.


\acknowledgments{Acknowledgments}

We are benefited a lot with discussions from Prof Cvetic during KITPC project around June 2014. Wei Xu would like to thank professor Geoffrey Comp\`ere, Hong L\"u and Jian-wei Mei for useful conversations. This work is partially supported by the Natural Science Foundation of China (NSFC) under Grant No.11075078 and by the project of knowledge innovation program of Chinese Academy of Sciences. Wei Xu was supported by the Research Innovation Fund of Huazhong University of Science and Technology (2014TS125).





\conflictofinterests{Conflicts of Interest}

The authors declare no conflict of interest.

\bibliographystyle{mdpi}
\makeatletter
\renewcommand\@biblabel[1]{#1. }
\makeatother

\providecommand{\href}[2]{#2}\begingroup
\footnotesize\itemsep=0pt
\providecommand{\eprint}[2][]{\href{http://arxiv.org/abs/#2}{arXiv:#2}}



%


%

\end{document}